\documentstyle[12pt]{article}

\def\title{\bgroup\obeylines\everypar={\hskip\parfillskip}\large
\bf\vrule height1cm width 0pt\relax} \def\endtitle{\vskip1sp\egroup}
\def\author#1{\hbox to\textwidth{\hss\vrule height.9cm width0pt\relax
#1\hss}} \def\contauthor#1{\hbox to\textwidth{\hss\vrule width0pt\relax
#1\hss}} \def\moreauthors#1{\hbox to\textwidth{\hss\vrule height.8cm
width0pt\relax #1\hss}}
\def\instit{\bgroup\small\it\obeylines\everypar{\hskip\parfillskip}}
\def\endinstit{\vskip1sp\egroup}

\newcommand{\be}{\begin{equation}}
\newcommand{\ee}{\end{equation}}
\newcommand{\bea}{\begin{eqnarray}}
\newcommand{\eea}{\end{eqnarray}}

\begin{document}

\vglue -4 true cm
\vskip -4 true cm

\begin{center}
{\hfill }{ FTUAM/92-39}
\end{center}

\vskip 1.5 true cm

\begin{title}
Instanton-like contributions to the dynamics
 of Yang-Mills fields on the twisted torus
\end{title}

\medskip
\begin{center}
 The RTN Collaboration\footnote{
Partially supported by DGA-PIT-2/89 and CICyT: TIC1161/90-E,
 TIC91-0963-C02,
AEN90-0034, AEN90-0272 and AEN90-0030.
}:

\medskip

M. Garc\'{\i}a P\'erez, A. Gonz\'alez-Arroyo and P. Mart\'{\i}nez,

{\it Departamento de F\'{\i}sica Te\'orica C-XI, Universidad Aut\'onoma
 de Madrid,
\\ 28049 Madrid, Spain,}

\smallskip

L. A. Fern\'andez, A. Mu\~noz Sudupe and J. J. Ruiz-Lorenzo,

{\it Departamento de F\'{\i}sica Te\'orica, Universidad Complutense
 de Madrid,
\\ 28040 Madrid, Spain,}

\smallskip

V. Azcoiti, I. Campos, J. C. Ciria, A. Cruz, D. I\~niguez,
F. Lesmes,\\ C. E. Piedrafita, A. Rivero, A. Taranc\'on and P.
T\'ellez,

{\it Departamento de F\'{\i}sica Te\'orica, Universidad de Zaragoza,
\\ 50009 Zaragoza, Spain,}

\smallskip

D. Badoni and J. Pech\footnote{and
Fyzik\'aln\'\i\ \'ustav \v Akademie, Praha, Czech Rep.}

{\it INFN-Sezione di Roma,\\ Roma, Italy.}
\end{center}

\begin{center} \today \end{center} \medskip

\begin{abstract} We study
$SU(2)$ lattice gauge theory in small volumes and with twist
$\vec{m}=(1,1,1)$.  We investigate the presence of the periodic
instantons of	$Q=\frac{1}{2}$ and determine their free energy and their
contribution to the splitting	of energy flux sectors
\mbox{$E(\vec{e}=(1,1,1))-E(\vec{e}=(0,0,0)) $.} \end{abstract}

\newpage

\section{Introduction}

Yang-Mills theory is an extremely rich Quantum Field Theory.  Despite
its simple formulation, it gives rise to a wide variety of subtle,
beautiful and complicated phenomena, such as asymptotic freedom,
confinement, dimensional transmutation and generation of a gap, finite
temperature phase transition, etc.  As in all QFT's, the structure of
the vacuum of the theory encompasses all of its properties.  Presumably,
the precise value of most of the observables of the theory such as the
string tension or the glueball spectra can only be obtained numerically
by means of Monte-Carlo simulations of the lattice-regularized version
of the theory.  However, it seems both desirable and feasible to have a
qualitative and semi-quantitative understanding of the structure of the
vacuum from which the mentioned numerical values arise.  In this respect
a good deal of work has been done in the last 20 years.

By now, the possible different phases of the theory are well-known and a
compelling numerical evidence shows that at $T=0$ the (4-dimensional)
theory is in the confined phase.  The corresponding vacuum structure has
been described as dual (electric $\leftrightarrow$ magnetic) to a
superconductor \cite{thooft1,mandel1}.  Despite the attractive features
of such description there is yet no equivalent to a dual BCS theory.  An
interesting attempt in this direction has recently been investigated
\cite{smit,lat93}.

Different classical configurations have been considered responsible for
different properties of the theory.  Most notably instantons
\cite{thooft4,gross}, merons \cite{gross2}, and monopoles
\cite{thooft5,nair}.  Other type of configurations ``fluxons"
\cite{thf,mackf} lie on the basis of the Copenhagen
\cite{domain1,domain2,domain3} description of the QCD vacuum.  All these
configurations are important, no doubt, but there is no general
consensus on the most relevant piece responsible for the property of
confinement.

In this paper we have investigated a new type of classical configuration
which is related to fluxons and for $SU(2)$ shares with merons the
property of being associated with lumps of topological charge $Q=1/2$.
These configurations are only known numerically \cite{us1,us2} and are
periodic in space.  They emerge quite naturally when considering gauge
fields on the torus with twisted boundary conditions.  Their role in the
QCD vacuum is at present unknown and in this paper we are taking the
first steps in the direction of their study.  Our results are restricted
to the study of their contribution to the path integral on relatively
small lattice toruses and for $\beta$ values where their behaviour would
turn out to be described by the semiclassical approximation.  The
prefactor, which has not been determined analytically, will be estimated
as a result of our computations.  Although this region is far from the
large volume one, where full Yang-Mills vacuum structure is recovered,
it can be seen as a second step in the description of the transition
from the small volume perturbative region to the large volume
confinement region.  It forms part of a general program initiated in
Ref. \cite{luscher} and \cite{koller2} for purely periodic boundary
conditions and in Ref. \cite{katony} for twisted boundary conditions.
For a review see Ref. \cite{vakron}.

At small volumes where, due to asymptotic freedom, perturbation theory
is a good approximation, the dynamics depends crucially of the boundary
conditions, and results obtained in the large $N$ limit \cite{TEK}
indicate that certain twisted boundary conditions are closer to the
infinite volume limit.  Since twist can be described as the presence of
${\bbb Z}_{N}$ magnetic flux in the torus, this fact agrees with the
Saviddy-Copenhagen picture of the vacuum \cite{zahed}

An important observable is given by the energy of a state carrying
non-zero ${\bbb Z}_2^{3}$ (for $SU(2)$) electric flux
\cite{fluxth1,fluxth2}.  On the torus this is a topologically conserved
quantum number described by a vector of integers modulo 2, \linebreak
$\vec{e}=(e_1,e_2,e_3)$.  Purely spatial Polyakov loops with winding
number $\vec{w}$ create electric flux $\vec{e}=\vec{w}\ ({\rm mod}\ 2)$.
Just as magnetic flux in ordinary superconductors, in the Yang-Mills
vacuum the minimum energy configuration in the presence of electric flux
is one where the flux is squeezed to a tube carrying a fixed energy
$\sigma$ per unit length.  To minimise energy, the tube must be a
straight line with winding number $\vec{e}$.  Due to the boundary
conditions, the tube cannot break and decay to the vacuum.  Thus we
expect \be E(\vec{e})-E(\vec{e}=0) = \ l_s \ \sigma \ \sqrt{|e_1|+|e_2|+
|e_3|}, \ee where $l_s$ is the length of the torus in each direction.

When the torus is small, the behaviour is quite different.  In the
absence of twist the levels are degenerate in perturbation theory
\cite{luscher} and the degeneracy is broken by tunnelling over a barrier
generated by quantum fluctuations \cite{koller2,koller1}.

In the case of the symmetric twist $\vec{m}=(1,1,1)$, there are two
spatial configurations (up to gauge transformations) which minimise the
potential energy.  Thus, in perturbation theory the eight-fold
degeneracy of the no-twist case is broken down to a two-fold degeneracy
between sectors with $\vec{e}$ and $\vec{e}+ \vec{m}$ electric fluxes
\cite{daniel}.  Polyakov loops which wind once along each spatial
direction will take opposite sign values in each of the classical minima
and hence we will refer to them as our order parameters $\Phi (t)$.

The splitting of sectors with $\vec{e}=(0,0,0)$ and $\vec{e}=(1,1,1)$
occurs beyond perturbation theory as a result from tunnelling between
both classical minima mediated by the twisted instanton configurations
of Ref. \cite{us1,us2}.  In the semiclassical approximation one gets

\begin{equation} \Delta E \equiv E(\vec{e}=(1,1,1))-E(\vec{e}=(0,0,0)) =
2 \ \frac{\langle \Delta N_I \rangle}{\Delta t}, \end{equation} where
the right-hand side is given by twice the mean number of instantons per
unit time.

In the rest of the paper we will show the results of our simulations,
leading to the computation of the energy splitting, which thus
complement the results of Ref. \cite{daniel} and serve to bridge the gap
between perturbation theory and the confinement phase.

\section{The Data} We have performed Monte Carlo simulations of the
Wilson-action $SU(2)$ Yang-Mills lattice theory with the spatial twist
$\vec{m}=(1,1,1)$ and no temporal twist, for various lattice sizes
$N_s^3\cdot N_t$ and values of $\beta$.  The simulations, which employed
the heat bath method, were performed on our reconfigurable 64-transputer
machine RTN \cite{RTN}.  The code was written in OCCAM language and the
parallelization algorithm consisted on distributing the lattice points
among the transputers and employing a checkerboard strategy: during the
updates, the transputers were organised in a ring and each of them
controlled a number of time slices.  With this method the
parallelization efficiency was very close to 95$\%$.  For some values of
$\beta$ the simulations were done in a single 8-transputers board of the
type contained in RTN and controlled by a PC.

One of our main concerns has been that of ensuring that our data are
appropriately thermalised and that two sequential measurements are
sufficiently uncorrelated.  As a matter of fact, we are dealing with the
dynamics of creation, annihilation and displacement of instantons which
extend over various lattice points.  The relaxation of these modes with
our local updating mechanism is much slower than for local quantities.
These autocorrelation times grow with $\beta$ and rather soon become
astronomically large.  For example at $\beta=3$ and $N_s=4$ a pair of
instantons do not annihilate after several hundred thousand sweeps,
while none is created if we start from a cold initial configuration with
no instantons.  This point was noticed some time ago by one of us (A.
G-A) and verified in Ref. \cite{stephenson}.  An overrelaxation
algorithm does not seem to improve things sizably in this case.
Fortunately, it turns out that for the values of $\beta$ which we have
used in this paper the autocorrelation times are still small compared
with the number of iterations.  To check this fact and ensure the
thermal character of our measurement sample, we have performed a number
of tests.  First of all we estimated the autocorrelation times by
analysing the evolution of the number of instantons from one measurement
to the next.  The number of iterations between subsequent measurements
$\Delta n$ was chosen to be larger than twice these autocorrelation
times.  In increasing order of $\beta$ values, $\Delta n$ was equal to
50, 250, 1000 and 3000 for $N_s = 4$, 1000 and 2000 for $N_s = 6$ and
1000 for $N_s = 8$.

As a final check that thermalization was attained, we have performed
simulations starting from different initial conditions and compared the
compatibility of the final results.  Some simulations started from one
of the cold configurations with no instantons, and others from hot ones
with many instantons.

As mentioned in the introduction, the main quantities which interest us
are the purely spatial Polyakov loops with winding number equal to one
around each of the three spatial directions, $\Phi$.  The corresponding
quantum operators carry electric flux $\vec{e}=(1,1,1)$ and produce
states with the same quantum numbers when acting on the vacuum.
Classically these loops take the values $\pm1$ on the two gauge
inequivalent classical vacua.  Instantons are configurations which
interpolate from one of the values to the other in time.  Thus,
operationally, we may identify the instanton time locations with those
of the sign flips of $\Phi$.

For large $\beta$ the choice of $\Phi$ is irrelevant.  The naive choice
$\Phi_0(t)$ is the spatial average of a Polyakov loop which is made of a
straight line in the \linebreak $x$-direction followed by one in the
$y$-direction and then one in the $z$-direction.  In Fig.~1a one can see
the structure of $\Phi_0(t)$ as a function of $t$ for a configuration
with $N_s=4$, $N_t=128$ and $\beta=15$.  The two-level structure and the
presence of the instantons is evident.

As $\beta$ decreases the thermal (quantum) fluctuations produce both a
decrease of $|\Phi_0|$ around each of the classical vacua together with
an increase in the fluctuations of $\Phi_0$ for each $t$.  The
combination of both effects make it very difficult for small $\beta$ to
separate between instantons and fluctuations around the classical vacua.
However, the problem is due mainly to very small wavelength fluctuations
and can be solved by the use of an improved order-parameter.  Such an
improved parameter can be obtained by averaging over various paths with
equal winding number and starting and ending at the same point.  The
average is performed by adding the path-dependent matrices themselves
and then normalizing the resulting matrix to have determinant equal to
1.  Finally the trace is taken and the result divided by 2.  It is clear
that in this way the value of the order parameter at the classical vacua
is still $\pm1$ provided the path-dependent matrices are taken with the
appropriate sign, while the uncorrelated fluctuations decrease like
$1/\sqrt{N(\gamma)}$ with $N(\gamma)$ the number of paths.  There are
various ways of selecting the set of paths.  In this paper we have
employed the fuzzying algorithm \cite{fuzz1,fuzz2}.  It amounts to
defining blocked links of length 2 as the $SU(2)$-projection of the sum
of purely spatial paths of length 2 and 4.  Then, applying the same
definition as $\Phi_0(t)$ for the resulting $(N_s/2)^3$ blocked lattice
and averaging over initial points, we get $\Phi_1(t)$.  In the case of
$N_s=4$ and $N_s=8$, we have applied a new blocking operation to the
previous one and obtained $\Phi_2(t)$.  For the case of $N_s=6$,
$\Phi_2(t)$ is defined by applying a scale 3 blocking transformation to
the configuration giving $\Phi_1(t)$.

In Fig.~1b we show how the use of the new order parameters improves the
situation of signal to background ratio.  For smaller $\beta$ and larger
$N_s$, even with the new operators, the fluctuations are non-negligible
and contaminate the sample.  Identifying the number and locations of the
instantons with that of the flips in sign of $\Phi$ would tend to
overestimate the former.  To extract the correct estimate of the number
of instantons we have employed two completely different approaches.  The
first one uses the different temporal distributions of instantons and
fluctuations to separate both samples on an statistical basis.  Later on
we will describe this approach in more detail.  The second approach is
based in reducing even further the amount of fluctuations by applying a
few cooling steps to the configurations \cite{cool1,cool2} and computing
$\Phi_0$, $\Phi_1$, $\Phi_2$ from the cooled configurations (Fig.~1c).
One might be worried by the possibility that cooling could affect the
instanton number and distribution.  Indeed, if a large number of cooling
steps were performed, one would end up in one of the configurations with
no instantons.  However, due to its local character, cooling washes away
faster short wavelength fluctuations while retaining the instanton
structure which is a local minimum of the action.  In our case we have
applied at most 3 cooling steps and feel confident that the instanton
structure has not been appreciably modified.  The validity of our final
results can be crosschecked by the comparison of both methods of
analysis which have completely different sources of error.

\section{Analysis of the data} Let us define the variable
$\sigma(t)={\rm sign} (\Phi(t))$ where $\Phi$ is any of the previously
mentioned order-parameters.  This is an Ising-like variable which is
defined on a one dimensional periodic lattice of length $N_t$.  We can
introduce a new Ising variable defined on links $\hat{\sigma}(t+1/2)$ as
$\sigma(t)\sigma(t+1)$.  In the absence of fluctuations, the presence of
$\hat{\sigma}$ taking the value $-1$ signals the appearance of an
instanton at the corresponding link.  Thus our most important result,
number of instantons per unit time, is given by \be \frac{1}{N_t}
\sum_{t} \ \left \langle \frac{1-\hat{\sigma}(t+1/2)}{2} \right \rangle
= \frac{\langle N_I \rangle }{N_t}. \ee The expected distribution of the
variables $\hat{\sigma}$ is Poisson-like, {\it i.e.} $\hat{\sigma}(l)$
and $\hat{\sigma}(l^{'})$ are statistically independent variables for
$l\neq l^{'}$.  Indeed, it deviates from Poisson due to the overall
constraint that the number of instantons must be even, and to the
discretization of time.  We have checked that the distribution is indeed
what it should by means of two tests.  The first test concerns the
distribution of $N_I$.  The expected behaviour is binomial \be P(N_I) =
{\cal N} \left ( \begin{array} {l} N_t \\ N_I \end {array}\right )
A^{N_I}, \label{eq:poisd} \ee where ${\cal N}$ is a normalization factor
and $A$ is related to the binomial parameter, $x$, by: $A= x/(1-x)$.
The data for $N_s = 4$, $\beta = 2.44$ and three cooling steps are shown
in Fig.~2a compared with the prediction of Formula (\ref{eq:poisd}).

A second test of the behaviour concerns the time distribution of
instantons: \be D(l) = \frac{1}{N_t} \sum_{t} \left \langle
\frac{(1-\hat{\sigma}(t+1/2))}{2} \ \frac{(1-\hat{\sigma}(t+l+1/2))}{2}
\right \rangle = \frac{\langle N_I \rangle^2}{N_t^2}. \ee Essentially
one is checking the $l$-independence of $D(l)$.  In Fig.~2b, we show the
plot of $D(l)$ for $N_s=4$, $\beta=2.44$.  For $l>2$ the distribution is
indeed flat, fitting the data to a constant gives a $\chi^2$ per degree
of freedom of 1.4.

It is clear that as $\beta $ gets small there are fluctuations in the
order parameter which induce flips $(\hat{\sigma} = -1)$ not due to
instantons.  Nonetheless these fluctuations have a small correlation
length, given by the inverse mass gap in the $\vec{e} =(0,0,0)$ sector.
For large $\beta$ this gap is $2\sqrt{2} \pi /N_s$.  The departure from
flatness in the plot for $D(l)$ can be attributed to this phenomenon.
In all cases this deviation is only appreciable for $l\le 3$.  The only
exception is our data point at $N_s=8$ which is also plotted in this
Figure.  Since cooling tends to eliminate fluctuations, it should also
give a flatter distribution, as can be easily seen in our example of
Fig~2b.

A simple model that incorporates fluctuations is to consider that \be
\sigma(t) = \lambda (t) u(t) \ee where $\lambda$ and $u$ are two new
Ising variables.  The former $\lambda(t)$ are a bunch of identical
independent variables with mean $\langle \lambda \rangle $.  The latter
is the Ising variable describing the instantons and sharing the
Poisson-like distribution mentioned previously.  Our physical parameter
is related to the mean value $\langle u(t) u(t+1)\rangle $ and should be
independent on the order parameter that has been used (blocked or
cooled).  On the contrary the value of $\langle
\sigma(t)\sigma(t+1)\rangle$ does depend on the order parameter, since
$\lambda$ does.  In order to extract the number of instantons from our
data we need two observables from which to eliminate $\langle
\lambda\rangle$ and $\langle N_I\rangle$.  We have taken $D(1)$ and
$\langle \hat{\sigma}(t+1/2)\rangle$ to obtain our final values of
$\langle N_I \rangle/N_t$.

To check the final results we have used an alternative method to
determine the number of instantons which is based on imposing a dead
time cut.  Fluctuations produce flips which are clustered, while
instantons are uncorrelated in time.  Thus, we will neglect all flips
that are within a certain time distance $\tau$ from another flip.  In
this way we end up with widely separated flips which are hence certainly
instantons.  To obtain the estimate of the mean number of instantons we
should simply correct for the missing ``phase space" induced by the cut.
The method mimics the procedure followed often by experimentalists.

To conclude this section we comment that the quoted errors are obtained
by partitioning the data into a few groups, measuring the dispersion of
the results between groups and taking the square root of this dispersion
divided by the number of groups.  We have checked the independence of
the errors with respect to	group size.

\section{Results and Conclusions}

Table 1 summarises our results.  The value of ${\cal A} (N_s)$ follows
from the formula \be \langle N_I\rangle = \frac{N_t}{2} \ {\cal A}(N_s)\
\beta^2 \	{\rm exp} \{-\frac{\beta}{4} S_L(N_s)\}, \label{eq:semicl} \ee
where the mean value of the number of instantons is obtained in an
$N_s^3\cdot N_t$ lattice and $S_L(N_s)$ is the instanton lattice action.
We have $S_L(4)= 37.4927$, $S_L(6)= 38.6376$ and $S_L(8)=39.0016 $ (as
$N_s$ grows the value quickly approaches $4\pi^2$).  For every value of
$\beta$, $N_s$ and initial configuration we give several determinations
of ${\cal A} (N_s)$ obtained through different observables.  The first
four columns come from estimating the mean number of instantons by
substracting from the number of flips in the order parameter, the
contributions of fluctuations, according to the prescription given in
the previous section.  The different values correspond to zero, one, two
and three cooling steps if available.  The results agree quite nicely
among themselves despite the simplicity of the model used to substract
fluctuations.  One can nevertheless consider that the numbers are
affected by, in addition to the quoted statistical error, a systematic
error due to the procedure used to substract fluctuations.  By the
observed dispersion of the different determinations with different
coolings, one can see that the size of this systematic error is of the
order of the statistical error.  An exception is the case of $N_s = 8 $,
owing perhaps to the fact that the flip correlation function (Fig.~2b)
deviates from flatness for up to 5 or 6 time slices, while the model
predicts flatness after one time slice.  We should nevertheless stress
once more that, despite the mentioned systematic errors, our model to
substract fluctuations is remarkably successful since the number of
flips before cooling is typically 2 to 3 times larger than the number of
instantons.

In the Table we have also shown other determinations.  The fifth and
sixth entries are obtained by counting the number of flips which are
separated from other flips more (or equal) than 3 and 5 time-slices
(dead-time) respectively.  One can determine the Poisson parameter from
these numbers without making any hypothesis on the nature of the
observed flip clustering.  The price to pay for this model independence
is an increase in the size of statistical errors.

Finally, the last column is an estimate obtained in terms of the
correlator $G(t) = \langle \Phi_2(0) \ \Phi_2(t) \rangle $ by the
formula \be 1 - 2 \frac{ \langle N_{I} \rangle }{N_t} =
\frac{G(t+1)}{G(t)}, \ee Consistent determinations are gotten from
different values of $t$ in the range 2 to 10.  Errors tend to increase
at large separations.  The quoted number is a weighted average of time
distances 2, 4, 6, and 8.  The error given is the smallest of all the
errors for fixed $t$.

In summary, good agreement is obtained between all model independent
determinations, yielding an estimate which is normally slightly smaller
than the ones obtained from our model of fluctuations.  Nonetheless the
difference is, except for the $N_s = 8 $ case, within one or two
standard deviations.  No difference is found if for our correlators and
dead-time estimates we use the cooled configurations.

Let us now compare the different rows of the table.  First of all, there
are entries corresponding to the same values of $N_s$ and $\beta$, but
obtained starting from different initial configurations.  The agreement
of these estimates within errors provides a check that our data are
appropriately thermalised.  Next, we compare the results obtained from
the different values of $\beta$ but the same value of $N_s$.  Notice,
that our results for ${\cal A} (N_s)$ are consistent with each other.
This fact means that our data correspond to a region in parameter space
where the semiclassical approximation holds and thus expression (7) is
well satisfied.  One could expect corrections of order $1/\beta$ to this
formula to occur but, due to the limited range of explored $\beta$
values, they do not show up.  From the different values of $\beta$ one
can give an average determination of the prefactor ${\cal A} (N_s)$:
\bea {\cal A}(4)&=& 2.013\ (19)\ 10^8,\nonumber \\ {\cal A}(6)&=& 7.43\
(15)\ 10^8, \label{eq:aparam}\\ {\cal A}(8)&=& 15.57\ (66)\ 10^8,
\nonumber \eea obtained from the last column of Table 1.  The numbers
could include small $1/\beta$ corrections and may then differ slightly
from the asymptotic value which would follow from a one loop computation
of fluctuations around the instanton.

Finally, we arrive to a comparison among lattice sizes.  The relation
among the different values of $N_s$ follows from scaling.  In order to
show how scaling is reproduced by our results, we have plotted our
estimate for $\Delta E \cdot l_s $.
This is a dimensionless quantity which on
the lattice is determined from our results as \be \Delta E \cdot l_s =
-\log \left (1-2\frac{\left \langle N_I \right \rangle}{N_t}\right )
\cdot N_s. \ee

The lattice quantities do of course depend on $N_s$ and $\beta$, but
scaling predicts they should depend on the single variable $l_s=N_s
a(\beta)$.  In Fig.~3 we have plotted $\Delta E \cdot l_s $ as a
function of $l_s$.  Since our data are still in a region where
asymptotic scaling is not valid, we have taken \be a(\beta)= 400\
\exp\{-\frac{\log2}{0.205} \beta\} \ \ {\rm fm}, \ee where the
dependence on $\beta$ is extracted from recent results in the same
$\beta$ range \cite{HELLER}.

One can see that our results scale quite well, given the small values of
$N_s$ involved.  In particular, data from $N_s = 4$ and $6$ at $l_s =
0.512$ fm are fairly close to each other and close to our data point at
$N_s=8$ with slightly smaller $l_s$.  The same is true for the region
around $\l_s = 0.42 $ fm.  The continuous lines are the prediction of
the semiclassical formula (\ref{eq:semicl}) with ${\cal A}(N_s)$ given
by (\ref{eq:aparam}).  They can be seen to describe pretty well our data
points: It seems that the scaling limit is approached from above, since
the $N_s = 4$ data lies above the other two curves.  The $N_s =6$ and
$N_s =8$ curves are perfectly consistent within errors.  The continuum
limit curve for $\Delta E \cdot l_s $ is therefore expected to follow
closely the shape of these curves.  They can be well approximated by an
$l_s^3$ dependence, lying in between the $l_s^\frac{11}{3}$ predicted by
the continuum semiclassical approximation and $l_s^2$ which follows for
a finite value of the string tension.  If we extrapolate the $l_s^3$
dependence to larger values of $l_s$, the curve goes through the data
points obtained by Stephenson \cite{stephenson} $2.72(56)$ at $l_s =
0.68$ fm and $20(6)$ at $l_s = 1.43$ fm.

To conclude let us summarise our results and mention some open problems.
We have studied the occurrence of the twisted instanton configurations
of Refs. \cite {us1,us2} in SU(2) Yang-Mills theory for volumes $N_s=
4,\ 6$ and $ 8 $ and $\beta$ values ranging from 2.38 up to 2.6.  Our
results show that the data in this region follow the semiclassical
formulas both in the average number of instantons as in the time
distribution and Poisson-like number distribution.  From our results an
estimate of the prefactor appearing in the semiclassical expression (7)
is obtained.  From these values one can determine the splitting $\Delta
E$ between the $\vec{e}=(1,1,1)$ and $\vec{e}=(0,0,0)$ electric flux
sectors (which is zero in perturbation theory) in the region $l_s \in
(0.341,0.512)$ fm.  As the size of the torus increases $\Delta E$ should
approach the confinement prediction \linebreak $\sqrt{3} \ \sigma l_s$.
{}From different considerations one expects that behaviour to set in for
\linebreak $l_s \approx 1 $ fm, where the dilute gas approximation has
already broken down and our methods of analysis are not-applicable.  It
would be very interesting to investigate how the transition to this
region is achieved and whether instantons are still present and
identifiable.  It is encouraging to discover that the extrapolated value
of $\Delta E/(l_s \cdot \sqrt{3})$ at 1 fm is $(0.41(2) \ {\rm
Gev}^{-1})^2$, not far from the infinite volume value.

\newpage

\section*{Table Captions}

The values of ${\cal A}(N_s)$ appearing in formula (\ref{eq:semicl}) are
shown for all our different runs with cold and hot initial
configurations.  The first four estimates are obtained by substracting
from the number of flips of the order parameter after c cooling steps,
the contribution of fluctuations (Section 3).  The eighth and ninth
columns come from our dead-time cuts of 3 and 5 respectively.  The last
column comes from formula (8).

\newpage

\section*{Figure Captions}

\begin{itemize}
\item{Fig. 1}

\begin{itemize} \item{a)} The order parameter $\Phi_0$ is plotted as a
function of lattice time $t$ for a configuration with $\beta=15$ and
$N_s=4$. \item{b)} The order parameters $\Phi_0$ (thin line) and
$\Phi_2$ (intermediate line) for $\beta=2.5$ and $N_s=4$. \item{c)} For
$\beta=2.5$ and $N_s=6$ we plot $\Phi_0$ (thin line), $\Phi_2$
(intermediate line) and $\Phi_2$ after 3 coolings (thick line).
\end{itemize}

\item{Fig. 2}

\begin{itemize} \item{a)} The histogram of number of instantons observed
in our run for $\beta=2.44$ after 3 coolings, compared with the
prediction for a binomial distribution. \item{b)} We display the flip
correlation function $D(l)$ for 3 configurations: $\beta=2.6$, $N_s=8$
(diamonds) and $\beta=2.44$, $N_s=4$ before (filled circles) and after 3
cooling steps (empty circles). \end{itemize}

\item{Fig.3}

The results of $\Delta E\cdot l_s$ as a function of the spatial lenght
$l_s$.  Circles, triangles and squares come from our data for $N_s=4,6$
and 8 respectively.  The curves are the predictions of the semiclassical
formula (\ref{eq:semicl}) with ${\cal A}$ given by Eq.
(\ref{eq:aparam}). \end{itemize}

\end{document}